\documentclass[10pt, conference, compsocconf]{IEEEtran}
\usepackage{times,amsmath,epsfig,multirow,tabularx,supertabular,graphics}
\ifCLASSINFOpdf
\else
\fi
\begin{document}
%
\title{A Community-Based Sampling Method Using DPL for Online Social Networks}




%
\author{\IEEEauthorblockN{Seok-Ho Yoon\IEEEauthorrefmark{1},
Ki-Nam Kim\IEEEauthorrefmark{1},
Sang-Wook Kim\IEEEauthorrefmark{1}, and
Sunju Park\IEEEauthorrefmark{2}}
\IEEEauthorblockA{\IEEEauthorrefmark{1}Department of Electronics and Computer Engineering\\
Hanyang University, 222 Wangsimni-ro, Seongdong-gu, Seoul 133-791, Korea
\\ Email: wook@hanyang.ac.kr}
\IEEEauthorblockA{\IEEEauthorrefmark{2}School of Business\\
Yonsei University, 50 Yonsei-ro, Seodaemun-gu, Seoul 120-749, Korea
\\ Email: boxenju@yonsei.ac.kr}}


\maketitle

\begin{abstract}
In this paper, we propose a new graph sampling method for online
social networks that achieves the following. First, a sample graph
should reflect the ratio between the number of nodes and the number
of edges of the original graph. Second, a sample graph should
reflect the topology of the original graph. Third, sample graphs
should be consistent with each other when they are sampled from the
same original graph. The proposed method employs two techniques:
hierarchical community extraction and densification power law. The
proposed method partitions the original graph into a set of
communities to preserve the topology of the original graph. It also
uses the densification power law which captures the ratio between
the number of nodes and the number of edges in online social
networks. In experiments, we use several real-world online social
networks, create sample graphs using the existing methods and ours,
and analyze the differences between the sample graph by each
sampling method and the original graph.

\end{abstract}


%
\IEEEpeerreviewmaketitle

\section{Introduction}
As the number of people who use online social networks grows, there
have been significant research interests on online social network
analysis \cite{Alb99}\cite{Fal99}\cite{Kum06}\cite{Yoo09}. It is
difficult and often impossible to analyze an online social network
with millions of nodes in its entirety \cite{Les05}\cite{Les06}.
What we need is a sampling method that creates sample graphs that
are representative of the original online social network
\cite{Les06}\cite{Hub08}\cite{Lee06}\cite{Rib10}.

A `good' sample graph should be able to produce analytical results
that would have been attained with the original graph. This means,
the properties of a sample graph should be similar to those of the
original graph. In this paper, we propose a new sampling method that
satisfies a list of graph properties a `good' sample graph should
aim for. Given an original graph and a sample size, the proposed
sampling method creates a sample graph with the properties quite
similar to the original graph.

Existing graph sampling methods are classified into three groups
\cite{Les06}: sampling methods by random node selection, sampling
methods by random edge selection, and sampling methods by
exploration. Sampling methods by random node selection create a
sample graph by selecting a set of nodes and then including the
edges connecting the selected nodes. Sampling methods by random edge
selection create a sample graph by selecting a set of edges and then
including the nodes connected to the selected edges. Sampling
methods by exploration create a sample graph by selecting a seed
node uniformly at random, exploring the neighbor nodes, selecting
the nodes explored and the edges connecting them, and continuing to
explore more nodes \cite{Les06}.

We claim that existing graph sampling methods fail to create samples
that retain the properties of an original graph. In the existing
graph sampling methods, the sample size is determined either by the
number of nodes or by the number of edges. The sampling methods by
random node selection or by exploration repeatedly select nodes
until satisfying the given number of nodes without regard to the
number of edges. The sampling methods by random edge selection, on
the other hand, repeatedly select edges until satisfying the given
number of edges without regard to the number of nodes. As a result,
a sample graph may have the node-edge ratio quite different from
that of the original graph since the sample size is determined by a
single basis. Furthermore, the sampling methods by exploration
create a sample graph that reflects the properties of only the part
of the original graph (near the seed node where sampling is taken
place). Finally, the random selection inherent employed in existing
graph sampling methods tends to generate `random' samples; the
properties of sample graphs tend to be inconsistent with one another
and with the original graph.

In this paper, we propose a new sampling method based on two key
concepts: hierarchical community extraction and Densification Power
Law (DPL). The hierarchical community extraction partitions the
original graph into a set of densely-connected sub-graphs (i.e.,
communities), and uses a dendrogram to represent the hierarchy
between the set of communities
\cite{Cla04}\cite{New04A}\cite{New04B}. A set of sample sub-graphs,
one for each community, is created by selecting nodes within the
community with the probability of selecting a node to be in
proportion to its degree and by selecting edges connecting them. The
final sample graph is created by merging the sample sub-graphs while
retaining the connectivity across the communities using the
dendrogram. The Densification Power Law (DPL) represents the ratio
between the number of nodes and the number of edges in real-world
social networks \cite{Les05}. When creating a sample sub-graph, we
use the ratio between the number of nodes and the number of edges
given by the DPL as a guideline.

By combining hierarchical community-based sampling and DPL, we
overcome the problems with existing sampling methods. First, since
the method uses the DPL, the sample graph created by the proposed
method reflects the ratio between the number of nodes and the number
of edges in both local and entire regions of the original graph.
Second, since the method uses the hierarchy community extraction,
the sample graph reflects the topology of the original graph well.
Third, since the method considers both the topology and the
node-edge ratio of the original graph, the properties of sample
graphs obtained from the same original graph tend to be consistent
with one another and with the original graph.

Through experiments on several diverse real-world online social
networks, we demonstrate the effectiveness of our sampling method.
As a performance metric, we use five well-known properties of a
social network: degree distribution, singular value distribution,
singular vector distribution, average clustering coefficient
distribution, and hop distribution. The difference between several
existing methods and ours is evaluated using K-S D-statistics
(Kolmogorov-Smirnov D-statistics) \cite{Les06}. The analyses show
that the properties of the sample graph by our method are the most
similar to those of the original graph.

The paper consists of the following. Section II introduces existing
sampling methods and points out the problems with the existing
methods. Section III describes the proposed method and presents the
detailed process. Section IV compares the performance of the
proposed methods with those of the existing methods through
experiments. Section V summarizes and concludes the paper.

\section{Related Work}

In this section, we review the existing sampling methods and point
out their problems.

\subsection{Existing Sampling Methods}

Leskovec and Faloutsos classified the sampling methods into three
groups: sampling methods by random node selection, sampling methods
by random edge selection, and sampling methods by exploration
\cite{Les06}. Figure 1 shows an example of the sampling methods by
random node selection. The nodes and edges with solid lines are the
ones selected for the sample graph, and those with dashed lines are
the ones not selected. The sample graph in Figure 1 is created by
selecting a set of nodes uniformly at random and then by selecting
all of the edges connecting the selected nodes. If the sample size
is given, the method selects nodes repeatedly until satisfying `the
number of nodes' to be selected for the sample.

Sampling methods based on random node selection differ in the way
nodes are selected. Random Node (RN) sampling selects a set of nodes
uniformly at random. The sample graphs created by RN are expected to
reflect the properties of the original graph, as samples are
selected from the entire population space. In Random Degree Node
(RDN) sampling, the probability of a node being selected is
proportional to its degree. In Random PageRank Node (RPN) sampling,
the probability of a node being selected is proportional to the
authority score computed by PageRank \cite{Pag98}. The idea behind
RDN and RPN is to increase the chance of including `important' nodes
in a sample graph. The nodes with many edges, the hubs, are the
important nodes in a social network and should be included in a
sample graph \cite{Bar03}.

\begin{figure}[h]
\centering
\includegraphics[width=4cm]{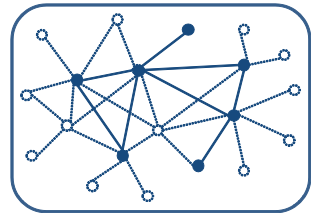}
\caption{Sampling methods by random node selection.} \label{fig_1}
\end{figure}

Figure 2 shows an example of the sampling methods by random edge
selection. The sample graph in the figure is created by selecting a
set of edges uniformly at random and then selecting all of the nodes
connected to the selected edges. If the sample size is given, the
method selects edges until satisfying `the number of edges'.

Similar to sampling methods by random node selection, sampling
methods by random edge selection differ in the way edges are
selected. Random Edge (RE) sampling selects a set of edges uniformly
at random and all the nodes connected to the selected edges. The
sample graphs created by RE tend not to reflect the structure of the
original graph since the high-degree nodes are selected more
frequently. Random Node Edge (RNE) sampling solves the problem by
selecting a node uniformly at random and then selects an edge
uniformly at random among the edges connected to the selected nodes
\cite{Les06}.

\begin{figure}[h]
\centering
\includegraphics[width=4cm]{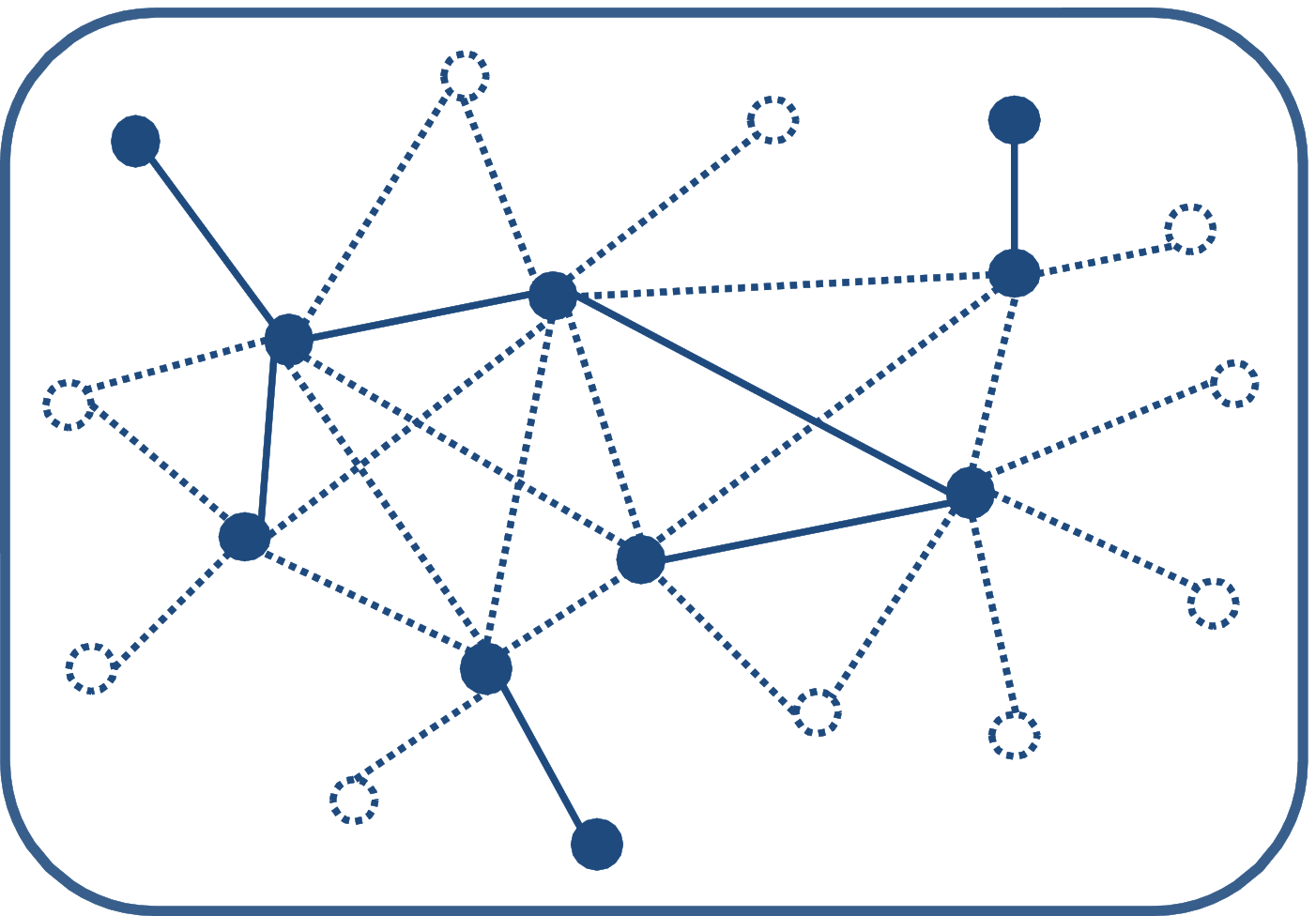}
\caption{Sampling methods by random edge selection.} \label{fig_2}
\end{figure}

The sampling methods by exploration create a sample graph by
selecting a seed node uniformly at random, exploring its neighbor
nodes, selecting all of the nodes explored and their connecting
edges, and continuing to explore more nodes. If the sample size is
given, the method selects nodes repeatedly from the original graph
until satisfying `the number of nodes'.

Depending on which edges to include in the sample, the exploration
methods are further classified into two: non-induced and induced.
The non-induced method includes only the edges explored in the
sample graph. The induced method includes not only the explored
edges but all of the edges connected to the selected nodes
\cite{Hub08}. Figure 3 shows the examples of sampling methods by
exploration. The node $S$ is a seed node. The method creates a
sample graph by exploring the nodes connected to the seed node, as
shown in Figure 3. \vspace{-0.2cm}
\begin{figure}[h]
\centering
\includegraphics[width=4cm]{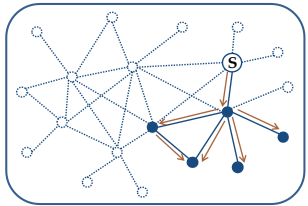}
\caption{Sampling methods by exploration.} \label{fig_3}
\end{figure}
\vspace{-0.2cm} Sampling methods by exploration include Random Walk
(RW) sampling, Random Jump (RJ) sampling, and Forest Fire (FF)
sampling. Both RW and RJ sampling methods use the concept of random
walk with restart \cite{Pag98}. The difference is the number of
seeds used. RW sampling uses a single seed node; RJ sampling uses a
set of seed nodes. Compared to RW and RJ that explores the graph
depth-first, FF sampling explores the graph breadth-first. FF
sampling picks a seed node at random, explores not a single but
multiple neighbor nodes. Then, it continues to explore the nodes
connected to the explored neighbor nodes recursively.

\subsection{Problems of Existing Sampling Methods}

We point out problems of the existing sampling methods by group.
First, sample graphs created by random node selection may have more
or fewer `edges' than the number of edges estimated by the ratio
between nodes and edges of the original graph, since the sampling
methods select nodes until satisfying the estimated number of nodes
without regard to the number of edges. For example, suppose that
sample size is 10\%, the number of nodes of original graph is 5,000,
and the number of edges of the graph is 10,000. The sample graph
should have 500 nodes and about 800 edges (In social networks the
number of edges tends to decrease exponentially  when the number of
nodes increases linearly. In Section III, we will explain it in
detail). Sampling methods by random node selection do not meet this
requirement.

Most random node selection methods select nodes at random. The
random selection tends to generate `random' samples; the properties
of the resulting graphs sampled from the same original graph tend to
be inconsistent with one another and with the original graph. RDN
and RPN do not select nodes at random but select nodes in proportion
to their degrees or authority scores. Sample graphs created by RDN
and RPN are more consistent, but they tend to be denser than the
original graph since they include many high-degree nodes.

Second, similar to the cases with random node selection, sample
graphs created by random edge selection may have more or fewer
`nodes' than the estimation provided by the ratio between nodes and
edges of the original graph. Also, the properties of the resulting
sample graphs created by random edge selection tend to be
inconsistent with one another and with the original graph since most
methods select edges at random.

Third, sampling methods by exploration repeatedly select nodes until
satisfying the estimated number of nodes, as done in random node
selection. Thus, similar to the cases with random node selection,
sampling by exploration may have more or fewer edges than the
estimation provided by the ratio between nodes and edges of the
original graph. The node-edge ratio of the sample graph created by
the non-induced method is closed to 1:1 since only the explored
nodes and edges are selected. The node-edge ratio of the sample
graph created by the induced method, on the other hand, is not 1:1
since the explored nodes and all of the edges connecting the
selected nodes are selected. When all of the edges connecting the
nodes in the connected graph are included into the sample graph as
in the induced methods, however, the sample graph would be denser
than the original graph. Also, the sample graphs created by
exploration are inconsistent since the methods select seed nodes at
random and the neighbor nodes of the seed node at random.

Furthermore, sample graphs created by exploration do not represent
entire graph but only the part near the seed node. RJ is regarded as
a solution to this problem. Sample graphs created by RJ reflect the
properties of the various quarters of the original graph better
because it uses multiple seeds.

\section{Proposed method}

In this section, we propose a new sampling method and describe its
process in detail.

\subsection{Overview}

In the previous section, we have pointed out the problems with
existing sampling methods. The new sampling method is designed to
achieve the following:

\vspace{0.2cm}
\begin{tabular}{|>{\centering}p{0.5cm}p{6.9cm}|}
    \hline

    (a)& The sample graph should reflect the node-edge ratio of each region of the original
    graph. \\
    (b)& The sample graph should reflect the node-edge ratio of the entire
original graph. \\ (c)& The sample graph should reflect the topology
of each region of the original graph.
\\ (d)& The sample graph should reflect the
topology of the entire graph.
\\(e)& The properties of graphs sampled from the same graph by the proposed
method should be consistent with one another.\\
\hline
\end{tabular}
\vspace{0.2cm}

The properties of regions of a social network may be different from
one another and also from those of the entire network. If we create
a sample graph which reflects the node-edge ratio of the entire
graph only, the node-edge ratio of a region of the sample graph may
not correctly represent its corresponding region of the original
graph. The properties of a graph are closely associated with the
topology of the graph. Thus, we should create a sample graph which
reflects both the topology of each region and the entire graph.
Finally, we should create sample graphs from the same graph whose
properties are consistent with one another and with the original
graph.

To create a sample graph that reflects the properties of the
original graph, the proposed sampling method utilizes two key
concepts: hierarchical community extraction and Densification Power
Law (DPL). First, it uses hierarchical community extraction to
partition the original graph into a set of densely-connected
sub-graph, i.e., communities. Hierarchical community extraction not
only partitions the original graph into a set of communities but
also creates a dendrogram that represents the partition hierarchy.
Figure 4 shows an example of the dendrogram. The large circle is the
community, and the edge among the circles represents the
parent-child relationship between communities. In Figure 4, a parent
community is partitioned into two children communities. After
partitioning the original graph into a set of communities, the
proposed method builds sample sub-graphs, one for each community.
Then, the method merges sub-graphs into a final sample graph from
bottom up, while taking the connections between the communities into
account using the dendrogram.

Second, the proposed method uses the DLP, when determining the
number of nodes and edges to be included in each sample sub-graph.
The DLP states that the number of nodes and the number of edges in a
social network follows the power law distribution \cite{Les05}.
Equation 1 shows DPL. In Equation 1, $e$ represents the number of
edges, and $n$ does the number of nodes. Typically, densification
exponent $\alpha$ takes the value between 1 and 2.

\vspace{-0.2cm}
\begin{figure}[h]
\centering
\includegraphics[width=8.5cm]{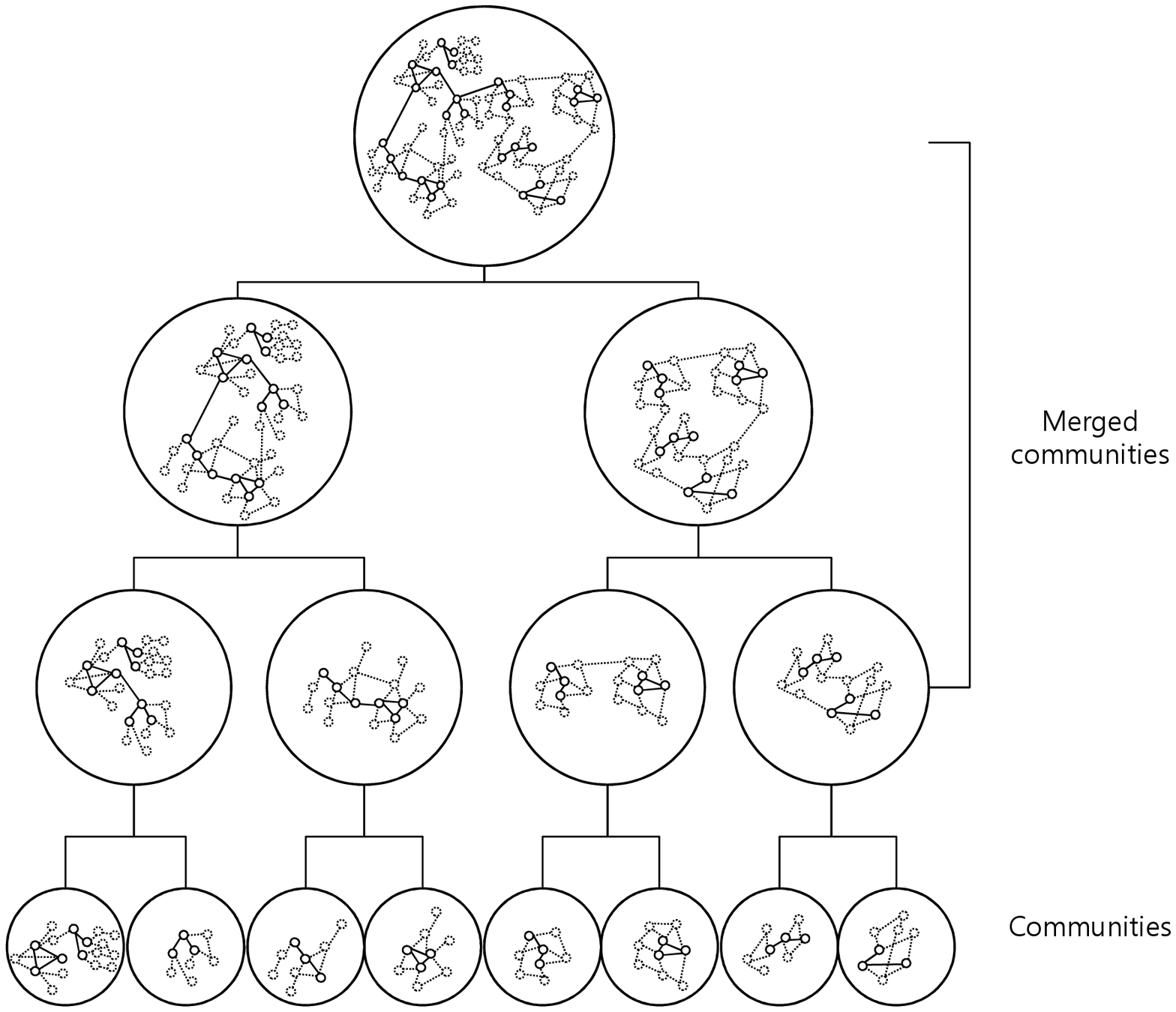}
\caption{A dendrogram and the process of the proposed method.}
\label{fig_4}
\end{figure}
\vspace{-0.2cm}
{\begin{align} &e \propto n^{\alpha} \label{eq:1}
\end{align}}
\vspace{-0.2cm}

In the existing sampling methods, the sample size is based on either
the number of nodes or the number of edges (but not both). In
comparison, the proposed method uses DPL when determining the size
of the sample graph. The proposed method estimates the number of
sample edges using DPL when the sample size is given as the number
of nodes in a social network. The proposed sampling method first
determines the value of $\alpha$ for each sub-graph (i.e.,
community) in the original graph based on the number of edges and
nodes within it. It determines the number of nodes to be included in
a sample community based on the number of nodes in its corresponding
community in the original graph. Then, it computes the number of
edges in the sample community by using the value of $\alpha$ for its
corresponding original community. A similar approach is used when
the sample size is given as the number of edges.

In summary, the proposed sampling method works as follows. First,
the method partitions the original graph into sub-graphs. Second, it
computes the densification exponent $\alpha$ based on the number of
nodes and edges for every community in the original graph. Third, it
builds sample sub-graphs by selecting nodes in proportion to their
degrees and edges connecting the selected nodes. The node-edge ratio
in a community is controlled by $\alpha$. Finally, it merges sample
sub-graphs into a final sample graph in a bottom up fashion, while
taking the connections between communities into account. For each
merged community, the node-edge ratio is also controlled by $\alpha$
of its corresponding community in the original graph. Figure 4 shows
the process of the proposed method. The nodes and edges in a
community with dashed lines are the ones not selected in the sample
graph, and those with solid lines are the ones selected in the
sample graph.

\subsection{Process of the proposed method}

In this section, we explain the process of the proposed method in
detail.

\subsubsection{Determining the number of sample nodes and sample edges}

The number of sample nodes and the number of sample edges in each
community are determined as follows. First, the proposed method
computes the number of sample nodes based on the sample size.
Second, based on the densification exponent $\alpha$, the method
computes the number of edges to be selected from each community.
This process makes sure that the sample graph reflects the node-edge
ratio in each community. For example, suppose the sample size is
10\%, the number of nodes and edges of in the original graph  are
500 and 1000, respectively. Based on equation (1), the densification
exponent $\alpha$ of the original graph is 1.11. The number of
sample nodes and sample edges from that community are 50 and 76,
respectively.

All the nodes in the original graph exist within the communities,
but some edges in the original graph exist between the communities.
Thus, we must determine the number of edges to be selected between
two communities (i.e., inter-community edges). The method determines
the number of edges to be selected between two child communities as
the difference between the number of edges to be selected from the
parent community and the number of edges to be selected from two
child communities. This provides the number of inter-community edges
to be selected between the two communities.

\subsubsection{Sampling with communities}

The proposed method creates a sample graph from each community as
follows. First, it selects nodes from each community until
satisfying the number of sample nodes pre-determined by the sample
size. Similar to RDN, the probability of selecting a node is in
proportion to its degree, which ensures important nodes are
selected. Second, it selects edges until satisfying the number of
edges determined by DPL. The probability of selecting an edge is
proportional to the sum of the degrees of the nodes connected to it.
Sample graphs created by this method are more consistent with one
another since the method selects high-degree nodes as in RDN. Also,
since the degree of nodes conveys topological information, each
sample sub-graph created by this method reflects the topology of
each sub-graph of the original graph.

Finally, the method creates the final sample graph by sampling edges
among communities in the reverse order of partition of the original
graph using hierarchical community extraction. There exist a few
edges that connect two communities. In social network analysis,
these few edges are defined as weak-tie \cite{Bar03}. The method
selects the inter-community edges in proportion to the sum of
degrees of nodes connected to it, similar to the way it selects
edges within the community.

The proposed method may use any hierarchical community extraction
method as long as it provides a dendrogram. We use the method which
automatically determines the number of communities to be extracted,
such as the modularity-based algorithm \cite{Cla04}\cite{New04B} and
cross-association (CA) \cite{Cha04}. Of course, one may use the
method which requires the number of communities as an input, such as
METIS \cite{Fri97} and chameleon \cite{Kar99}, as long as domain
experts supply the optimal number of communities. In this paper, we
have used the modularity-based algorithm, a well-known hierarchical
community extraction method \cite{Cla04}\cite{New04B}.

\section{Experiments}

In this section, we demonstrate the effectiveness of the proposed
method by comparing it with several existing sampling methods.

\subsection{Experimental Setup}

We use seven real-world online social networks in our experiments
\cite{Les10A}\cite{Les10B}\cite{Ric03}\cite{Les07}\cite{Kri07}.
First, `Wiki-vote' is a dataset collected from Wikipedia from the
day when the service opened to Jan 2008. A node represents a user of
Wikipedia, and an edge represents a recommendation between the
users. Second, `Email Enron' is a collection of emails from Enron. A
node represents an email address, and an edge represents a
communication between email addresses. Third, `Epinions' is a
dataset collected from epinions.com, a product review website. A
node represents a user, and an edge represents a recommendation
between the users. Fourth and fifth, `Hep\_ph' (High Energy
Physics-Phenomenology) and `Hep\_th' (High Energy Physics-Theory)
are the datasets from Arxiv website, a website that collected
unpublished papers, from Jan 1992 to Apr 2003. In both datasets, a
node represents a paper, and an edge represents a reference between
papers. Sixth, `AS' is a dataset collected from the log analysis of
border gateway protocol between the Autonomous Systems, from Nov
1997 to Jan 2000. A graph of routers comprising the Internet can be
organized into sub-graphs called Atonomous System (AS). A node
represents an AS, and an edge represents a communication between
ASs. Seventh, `Oregon' is a log data collected from Oregon routers,
from Mar 2001 to May 2001. A node represents a router in Oregon, and
an edge represents a communication between the routers. We generate
undirected graphs with these data. Table 1 shows the numbers of
nodes and edges in each dataset.

\begin{table}[h]
\centering \caption{The size of the dataset used in the experiments}
\begin{tabular}{|c|c|c|} \hline
&\# of nodes & \# of edges \\
\hline
Wiki-vote & 7,115 & 201,524\\
\hline
Email Enron & 36,692 & 367,662\\
\hline
Epinions & 75,879 & 811,480\\
\hline
Hep\_ph & 34,545 & 841,754\\
\hline
Hep\_th & 27,768 & 704,570\\
\hline
AS & 6,743 & 25,144\\
\hline
Oregon & 10,669 & 44,004\\
\hline
\end{tabular}
\end{table}

As performance metric, we use a set of five well-known graph
properties: degree distribution, singular value distribution,
singular vector distribution, average clustering coefficient (CC)
distribution and hop distribution \cite{Les06}. The degree
distribution is the distribution of the number of nodes with degree
$d$ for every degree $d$. The degree distribution of an social
network typically follows a power law distribution \cite{Fal99}. The
singular value distribution and the singular vector distribution are
the distributions computed by singular value decomposition (SVD) of
the graph adjacency matrix \cite{Kor97}. These two properties
represent the characteristics of the community structure of a graph.
The average CC distribution is the distribution of the average CC of
nodes for every degree $d$. The CC of a node is the ratio between
the number of edges among the node and neighbor nodes and the number
of possible edges among the node and neighbor nodes. If the number
of neighbors of a given node is $k$, the number of possible edges is
$\frac{(k(k-1))}{2}$. The hop is the minimum distance between two
nodes. The hop distribution is the distribution of the number of
reachable pairs in hop $h$ for every hop $h$ \cite{Fal99}.

The sampling methods compared in our experiments are the proposed
method, RN, RDN, RPN, RE, RNE, RW, RJ, and FF. We use both the
non-induced versions of RW, RJ, and FF, and the induced versions of
RW (RW(i)), RJ (RJ(i)), and FF (FF(i)), respectively. The proposed
method is denoted as `C+D' (Community + DPL). We set the probability
of restart to be 0.15 for RW and RJ. For FF, we set $P_{f}$ to be
0.3, the best value suggested in \cite{Les06}. In all methods, we
use the sample size of 10\%.

\subsection{Performance comparisons}

In this section, we compare the performance of the proposed method
with other sampling methods. First, we visually inspect the CC and
degree distributions. Second, we examine the performance of various
sampling methods more rigorously using K-S D-statistics. Third, we
check the consistency among the graphs sampled from the same
original graph by each sampling method. Finally, we examine the
densification exponent of each sampling method.

\subsubsection{CC and degree distributions}

We examine how well the sample graph reflects the properties of the
original graph. The evaluation is based on visual inspection on the
CC distributions and degree distributions of the sample graphs by
various sampling methods. The comparison between the other
distributions of the original graph and those of the sampling
methods are not presented since they show no visible difference.

Figure 5 depicts the CC distribution of the original graph and that
of each sampling method. In Figure 5, the x-axis represents the
degree of nodes, and the y-axis represents the CC of nodes. The CC
distribution of the original graph is similar to that of the sample
graph created by the proposed method. The CC distributions of the
sample graphs created by RDN, induced FF, and induced RW are also
similar to that of the original graph. The CC distributions of the
sample graphs created by RN and RE, however, are quite different
from that of the original graph.

\begin{figure}[h]
\centering
\includegraphics[width=2.5in]{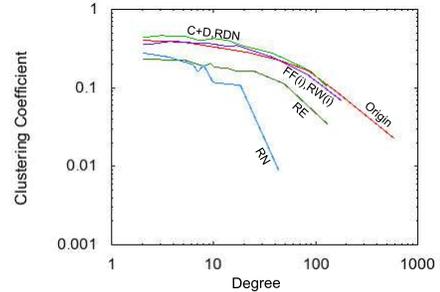}
\caption{CC distributions of the original graph and the sample
graphs.} \label{fig_5}
\end{figure}

Figure 6 shows the degree distribution of the original graph and the
sample graphs created by various sampling methods. In Figure 6, the
x-axis represents the degree of nodes, and the y-axis represents the
number of nodes. Again, the shape of the degree distribution of the
original graph is similar to that of the sample graph created by the
proposed method. The degree distribution of the sample graph created
by RN is similar to that of the original graph, but those by the
other methods are quite different from that of the original graph.
The visual inspection suggests that the propose method creates a
sample graph with the properties most similar to those of the
original graph.

\begin{figure}[h]
\centering
\includegraphics[width=2.5in]{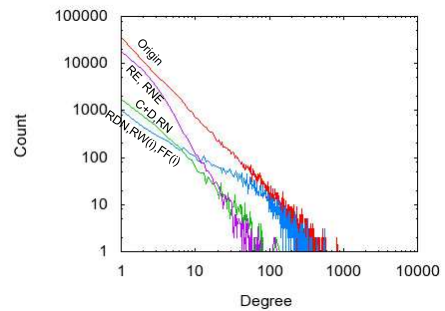}
\caption{Degree distributions of the original graphs and the sample
graphs.} \label{fig_6}
\end{figure}

\subsubsection{K-S D-statistics}

We compute the difference between the five properties of the
original graph and those of the sample graph using K-S D-statistics
(Kolmogorov-Smirnov D-statistics). K-S D-statistics computes the
maximum difference between the cumulative distribution function of
the original graph and that of the sample graph (see Equation (2)).
The $D$ value computed by K-S D-statistics is between 0 and 1. As
the value approaches 0, the property of the sample graph is more
similar to that of the original graph.

{\begin{align} &D=max_{x}(|F(x)-F'(x)|) \label{eq:2}
\end{align}}

The size of a sample graph matters; The closer the size of a sample
graph is to that of the original graph, the more the properties of
the sample graph are similar to those of the original graph
\cite{Les06}. For fair comparison, therefore, we should use sample
graphs with the same number of nodes or edges. However, the existing
sampling methods determine the size of a sample graph using either
the number of nodes or the number of edges. Thus, node-based methods
and edge-based methods could not be compared fairly since the number
of nodes and edges of the sample graphs created by each sampling
method could not be standardized. The proposed method can be
compared fairly with both node-based and edge-based sampling
methods. Thus, we separate the comparisons into two groups: the
comparison between the proposed method and node-based sampling
methods and the comparison between the proposed method and
edge-based sampling methods. We create 10 sample graphs by each
sampling method and then compute the average of the D-statistics of
the sample graphs.

Tables 2 and 3 show the comparison of the proposed method and
node-based sampling methods and the comparison of the proposed
method and edge-based sampling methods, respectively. In Tables 2
and 3, the numbers between 0 and 1 represent the D-statistics for
five different properties, and $R$ represents the ranking of the
sampling methods. $Avg$ represents the average of the five
D-statistics.

As shown in Tables 2 and 3, the performance of the proposed method
is consistently higher than those of existing sampling methods. In
Table 2, the proposed method is ranked high in all properties except
hop distribution. In degree distribution and singular value
distribution properties, where the proposed method is ranked the
first, the difference between the D-statistic values of the top and
the second is quite significant. In comparison, the D-statistics
values of hop distribution, where the proposed method is ranked the
sixth, all of the sampling methods are quite similar. Thus, the
sample graph created by the proposed method reflects the properties
of the original graph the best. In particular, the sample graph
created by the proposed method reflects the degree, singular value,
and singular vector distributions of the original graph well. This
is because the proposed method creates a sample graph by selecting
nodes and edges in proportion to their degrees and using the
hierarchical community extraction. Note that the D-statistics value
of CC of non-induced FF is 1. Since non-induced FF selects only the
explored nodes and edges and does not explore the explored nodes
again, the sample graph created by non-induced FF is a tree, which
results in a significant difference in the CC distribution of
non-induced FF and that of the original graph. In Table 3, the
proposed method is ranked the first in all properties.

In Tables 2 and 3, the variance of D-statistics in each property
differs. When comparing the performance of the sampling methods
using the average D-statistics of five properties, the overall
ranking depends on the D-statistics with wide variation. To avoid
this problem, we also compare the performance of the sampling
methods by normalizing the D-statistics. We normalize the
D-statistics values of each property by min-max normalization
 \cite{Han06}. The numbers in parentheses in Tables 2 and 3 refer to
normalized values and ranks computed with the normalized values. The
proposed method outperforms the other sampling methods. The
difference between the average score of the top and the second with
normalization is larger than that obtained without normalization.

\begin{table*}
\centering \caption{The proposed method vs. node-based sampling
methods}
\begin{tabular}{|c|c|c|c|c|c|c|c|c|c|c|c|c|} \hline
&Degree&R&Sval&R&Svec&R&CC&R&Hop&R&Avg&R \\
\hline
C+D & 0.132 & 1 & 0.044 & 1 & 0.176 & 2 & 0.338 & 3 & 0.045 & 6 & 0.147 (0.041) & 1 (1)\\
\hline

RN & 0.229 & 2 & 0.138 & 8 & 0.211 & 5 & 0.402 & 7 & 0.047 & 7 & 0.205 (0.257) & 6 (6)\\
\hline

RDN & 0.258 & 5 & 0.085 & 3 & 0.215 & 6 & 0.367 & 4 & 0.032 & 2 & 0.191 (0.172) & 2 (3)\\
\hline

RPN & 0.256 & 4 & 0.091 & 5 & 0.175 & 1 & 0.379 & 5 & 0.027 & 1 & 0.186 (0.166) & 3 (2)\\
\hline

RW & 0.229 & 2 & 0.128 & 7 & 0.189 & 4 & 0.470 & 8 & 0.052 & 8 & 0.214 (0.260) & 7 (7)\\
\hline

RJ & 0.353 & 9 & 0.183 & 10 & 0.436 & 9 & 0.592 & 9 & 0.044 & 5 & 0.322 (0.502) & 9 (9)\\
\hline

FF & 0.309 & 8 & 0.142 & 9 & 0.860 & 10 & 1.000 & 10 & 0.215 & 10 & 0.505 (0.836) & 10 (10)\\
\hline

RW(i) & 0.293 & 6 & 0.088 & 4 & 0.223 & 7 & 0.335 & 2 & 0.036 & 3 & 0.195 (0.193) & 5 (4)\\
\hline

RJ(i) & 0.506 & 10 & 0.104 & 6 & 0.185 & 3 & 0.380 & 6 & 0.036 & 4 & 0.242 (0.381) & 8 (8)\\
\hline

FF(i) & 0.305 & 7 & 0.084 & 2 & 0.265 & 8 & 0.260 & 1 & 0.058 & 9 & 0.195 (0.210) & 4 (5)\\
\hline
\end{tabular}
\end{table*}

\begin{table*}
\centering \caption{The proposed method vs. edge-based sampling
methods}
\begin{tabular}{|c|c|c|c|c|c|c|c|c|c|c|c|c|} \hline
&Degree&R&Sval&R&Svec&R&CC&R&Hop&R&Avg&R \\
\hline
C+D & 0.128 & 1 & 0.041 & 1 & 0.187 & 1 & 0.303 & 1 & 0.042 & 1 & 0.140 (0) & 1 (1)\\
\hline

RE & 0.258 & 2 & 0.130 & 2 & 0.349 & 2 & 0.360 & 2 & 0.054 & 2 & 0.230 (0.586) & 2 (2)\\
\hline

RNE & 0.302 & 3 & 0.143 & 3 & 0.580 & 3 & 0.518 & 3 & 0.061 & 3 & 0.320 (1) & 3 (3)\\
\hline

\end{tabular}
\end{table*}

\subsubsection{Consistency of sample graphs}

In this set of experiments, we evaluate the consistency of various
sampling methods. The consistency is measured by the standard
deviation of D-statistics of sample graphs. As shown in Table 4, the
standard deviation of RDN is the lowest, followed by RPN, the
proposed method, and RE. Note that the average standard deviations
of lower-ranked sampling methods are quite high. Since RDN and RPN
select high-degree nodes, the sample graphs tend to be consistent.
RE also selects many high-degree nodes, because it includes the
nodes connected to selected edges into the sample graph though edges
are selected at random. Thus, the sample graphs created by RE tend
to be consistent. The sample graphs created by the proposed method
are consistent since it selects nodes and edges in proportion to
their degree like RDN and RPN. The sample graphs created by the
proposed method is somewhat less consistent than those of RDN and
RPN, because RDN and RPN select all of the edges connected to
selected nodes while the proposed method selects edges in proportion
to degree of nodes connecting the edges. The proposed method
generates consistent sample graphs and ranked the third among
various sampling methods.

\begin{table*}
\centering \caption{Standard deviation of the sample graphs created
by various sampling methods}
\begin{tabular}{|c|c|c|c|c|c|c|c|c|c|c|c|c|} \hline
&Degree&R&Sval&R&Svec&R&CC&R&Hop&R&Avg&R \\
\hline
C+D & 0.0002 & 3 & 0.0001 & 7 & 0.0021 & 7 & 0.0005 & 3 & 0.0004 & 6 & 0.0006 & 3\\
\hline

RN & 0.0053 & 10 & 0.0050 & 11 & 0.0093 & 9 & 0.0230 & 12 & 0.0006 & 10 & 0.086 & 10\\
\hline

RDN & 0.0002 & 2 & 0 & 3 & 0.0002 & 1 & 0.0003 & 2 & 0.0002 & 1 & 0.0002 & 1\\
\hline

RPN & 0.0003 & 4 & 0 & 5 & 0.0002 & 2 & 0.0006 & 4 & 0.0003 & 3 & 0.0003 & 2\\
\hline

RE & 0 & 1 & 0 & 2 & 0.0003 & 3 & 0.0020 & 6 & 0.0004 & 5 & 0.0006 & 3\\
\hline

RNE & 0.0004 & 6 & 0 & 1 & 0.0113 & 10 & 0.0026 & 7 & 0.0004 & 7 & 0.0029 & 8\\
\hline

RW & 0.0072 & 11 & 0.0007 & 9 & 0.0056 & 8 & 0.0207 & 11 & 0.0005 & 9 & 0.0069 & 9\\
\hline

RJ & 0.0004 & 5 & 0.0001 & 6 & 0.1020 & 11 & 0.0109 & 10 & 0.0003 & 4 & 0.0228 & 11\\
\hline

FF & 0.0092 & 12 & 0.0125 & 12 & 0.8637 & 12 & 0 & 1 & 0.0048 & 12 & 0.1780 & 12\\
\hline

RW(i) & 0.0024 & 9 & 0.0005 & 8 & 0.0014 & 4 & 0.0070 & 9 & 0.0005 & 8 & 0.0024 & 6\\
\hline

RJ(i) & 0.0005 & 7 & 0 & 4 & 0.0018 & 6 & 0.0008 & 5 & 0.0002 & 2 & 0.0007 & 5\\
\hline

FF(i) & 0.0020 & 8 & 0.0015 & 10 & 0.0016 & 5 & 0.0068 & 8 & 0.0008 & 11 & 0.0025 & 7\\
\hline

\end{tabular}
\end{table*}

\subsubsection{Densification exponent}

The sampling target in this paper is a social network. Thus, the
ratio of the number of nodes and the number of edges in the sample
graph should follow the DPL. In this set of experiments, we examine
whether the sample graphs reflect the node-edge ratio of the
original graph. Table 5 shows the difference between the
densification exponent of the sample graph created by each sampling
method and that of the original graph.

\vspace{-0.2cm}
\begin{table}[h]
\centering \caption{Difference between the densification exponent of
the sample graph and that of the original graph}
\begin{tabular}{|c|c|} \hline
Sampling method & difference \\
\hline C+D & 0.033 \\\hline
 RN & -0.063 \\\hline
 RDN & 0.135 \\\hline
 RPN & 0.104 \\\hline
 RE & -0.065 \\\hline
 RNE & -0.162 \\\hline
 RW & -0.096 \\\hline
 RJ & -0.142 \\\hline
 FF & -0.156 \\\hline
 RW(i) &  0.141 \\\hline
RJ(i) & 0.105 \\\hline
 FF(i) & 0.149 \\\hline
\end{tabular}
\end{table}

\vspace{-0.2cm}

The sample graph created by the proposed method reflects the
node-edge ratio of the original graph more than the others. The
result is not surprising since the proposed method takes into
account the node-edge ratio of the original graph when creating a
sample graph. Of course, the node-edge ratio of the sample graph
created by the proposed method is not exactly the same as that of
the original graph. It is because the proposed method uses not the
node-edge ratio of the entire graph but the node-edge ratio of each
partitioned sub-graph.

The node-edge ratios of the sample graphs created by RN and RE are
slightly lower than that of the original graph. In the case of RN,
because sample nodes are selected uniformly at random from the
entire population space, the selected nodes often do not have edges
between them. Similarly, in the case of RE, because sample edges are
selected uniformly at random from the entire population, the nodes
connected to the selected edges often form an island. In both cases,
RN and RE end up with a sample graph sparser than the original
graph. RNE is the lowest, because RNE  tries to avoid the problem
that RE selects many high-degree nodes. In contrast, the node-edge
ratio of the sample graphs created by RDN and RPN is much higher
than that of the original graph because RDN and RPN select many
high-degree nodes.

The sample graphs created by non-induced RW, RJ, and FF are much
lower than that of the original graph because the method selects
only explored nodes and edges. In contrast, the sample graphs
created by induced RW, RJ, and FF are denser than that of the
original graph because they select not only explored nodes and edges
but also all of edges connecting selected nodes. RJ uses multiple
seed nodes while RW and FF use a single seed node. Thus, the
node-edge ratio of the sample graph created by RJ is lower than
those of RW and RJ even though that of the sample graph created by
RJ is denser than that of the original graph

\subsection{The effectiveness of the techniques used in the proposed method}

In the previous section, we have shown that the proposed method
outperforms the other sampling methods. As the proposed method is
based on two techniques, community-based sampling and DPL-based
sampling, we examine the effectiveness of these two techniques.

\subsubsection{The effectiveness of community-based sampling}

First, we examine the effectiveness of the community-based sampling
technique when creating a sample graph that reflects the properties
of the original graph. We apply the community-based sampling
technique to the existing sampling methods and evaluate the
performance of the sampling methods with community-based sampling.

For experiments, four representative sampling methods are selected:
RN for sampling by random node selection, RE for sampling by random
edge selection, RW for sampling by exploration, and RDN whose
selection process is similar to that of the proposed method. We call
RN, RE, RW, and RDN with community-based sampling as community-based
RN, community-based RE, community-based RW, and community-based RDN,
respectively. To apply the community-based method to the existing
sampling methods, we have done the following. The community-based RN
selects nodes at random from each community partitioned by
hierarchical community-extraction method and then selects all of
edges connecting the selected nodes. The community-based RE selects
edges at random from each community and then selects all of nodes
connected to the selected edges. The community-based RDN selects
nodes in proportion to the degree of a node from each community and
then selects all of edges connecting the selected nodes. The
community-based RW selects a seed node from each community and then
selects all of nodes explored by exploring neighbors of the seed
node. It selects all of edges connecting selected nodes. The
experimental setup and method are the same as those in Section
IV.B.2.

Table 6 compares the performance of the existing methods with and
without the application of community-based sampling. In Table 6,
`CBased' represents a method with community-based sampling. Table 6
confirms that the community-based sampling methods outperform the
original sampling methods.

Community-based sampling makes the sample graph to reflect the
topology of the original graph better. For example, the nodes in a
community are densely connected to each other. Thus, if
community-based RN selects nodes and edges in a community, the nodes
in the sample graph can be thought densely connected to each other.
Thus, community-based RN creates a sample graph that reflects the
properties of the original graph better than RN. The performance of
all community-based sampling methods improves for a similar reason.
We conclude community-based sampling is an effective technique for
creating a sample graph which reflects the properties of the
original graph.

\begin{table}[h]
\centering \caption{Comparison of the sampling methods with and
without community-based sampling}
\begin{tabular}{|c|c|c|c|c|c|c|} \hline
&Degree&Sval&Svec&CC&Hop&Avg \\
\hline\hline
RN & 0.289 & 0.138 & 0.211 & 0.402 & 0.047 & 0.217\\
\hline

CBasedRN & 0.273 & 0.112 & 0.207 & 0.372 & 0.052 & 0.203\\
\hline\hline

RDN & 0.549 & 0.085 & 0.215 & 0.367 & 0.032 & 0.250\\
\hline

CBasedRDN & 0.537 & 0.085 & 0.198 & 0.384 & 0.035 & 0.248\\
\hline\hline

RE & 0.210 & 0.130 & 0.349 & 0.360 & 0.054 & 0.220\\
\hline

CBasedRE & 0.188 & 0.118 & 0.339 & 0.359 & 0.031 & 0.207\\
\hline\hline

RW & 0.540 & 0.128 & 0.189 & 0.470 & 0.052 & 0.269\\
\hline

CBasedRW & 0.487 & 0.090 & 0.188 & 0.321 & 0.037 & 0.225\\
\hline

\end{tabular}
\end{table}

\subsubsection{The effectiveness of DPL-based sampling}
In this section, we examine the effectiveness of the DPL-based
sampling technique for creating a sample graph that reflects the
properties of the original graph. We apply the DPL-based sampling
technique to the existing sampling methods and evaluate the
performance of the sampling methods with and without DPL-based
sampling.

Similar to the previous experiments, we use RN, RE, RW, and RDN as
representative sampling methods. We should make sure the node-edge
ratio of the sample graph is the same as that of the original graph.
A simple way to achieve this would be to  create a sample graph by
the original method and then include or remove edges to match the
node-edge ratio. Note that the inclusion or removal of nodes would
not make the sample graph with the desired node-edge ratio because
when nodes are removed, edges connecting the removed nodes are
removed automatically. Thus, we  insert or remove only `edges' from
the sample graph created by the existing method. The number of edges
in the sample graph created by RN, RE, and RW are less than the
number of edges computed by the node-edge ratio of the original
graph. Thus, DPL-based RN, RE, and RW include more edges in
proportion to the sum of degrees of two nodes connected to the
edges. DPL-based RDN retains the edges in proportion to the sum of
degrees of two nodes connected to the edges and removes the rest.
The experimental setup and method are equal to those of Section
IV.B.2.

Table 7 compares the existing sampling methods with and without
DPL-based sampling. In Table 7, `DBased' represents a method with
DPL-based sampling. Compare to Table 6, Table 7 shows that not all
of the DPL-based sampling methods outperform the original sampling
methods. DPL-based RW and DPL-based RDN are better than original RW
and RDN, respectively, while original RN and RE are better than
DPL-based RN and DPL-based RE, respectively. From these results, one
may conclude the DPL-based sampling technique is not effective. This
conclusion, however, is incorrect. community-based RDN can be viewed
as the proposed method without DPL-based sampling. When we compare
the performance of community-based RDN (in Table 6) and that of the
proposed method (in Table 2), the proposed method is better than
community-based RDN, which indicates the DPL-based sampling
technique is effective in the proposed method. A more correct
interpretation of the results in Table 7 is that it is difficult to
apply DPL-based sampling to those methods. The simple inclusion or
removal of edges to the sample graph created by the existing methods
fails to keep the key concept of each sampling method.

\begin{table}[h]
\centering \caption{Comparison of the sampling methods with and
without DPL-based sampling}
\begin{tabular}{|c|c|c|c|c|c|c|} \hline
&Degree&Sval&Svec&CC&Hop&Avg \\
\hline\hline
RN & 0.289 & 0.138 & 0.211 & 0.402 & 0.047 & 0.217\\
\hline

DBasedRN & 0.430 & 0.160 & 0.355 & 0.478 & 0.215 & 0.328\\
\hline\hline

RDN & 0.549 & 0.085 & 0.215 & 0.367 & 0.032 & 0.250\\
\hline

DBasedRDN & 0.343 & 0.312 & 0.267 & 0.266 & 0.028 & 0.243\\
\hline\hline

RE & 0.210 & 0.130 & 0.349 & 0.360 & 0.054 & 0.220\\
\hline

DBasedRE & 0.395 & 0.062 & 0.526 & 0.717 & 0.258 & 0.392\\
\hline\hline

RW & 0.540 & 0.128 & 0.189 & 0.470 & 0.052 & 0.269\\
\hline

DBasedRW & 0.348 & 0.290 & 0.259 & 0.230 & 0.028 & 0.231\\
\hline
\end{tabular}
\end{table}

\subsubsection{The performance of the proposed method with different densification exponent}

We have not been able to conclude from Table 7 that the DPL-based
sampling technique is effective. In this section, we show the
effectiveness of the DPL-based sampling technique by comparing the
performance of the proposed method with different densification
exponent.

In Table 8, $d_{\alpha}$ represents the difference between the
densification exponents of the original graph and the sample graph.
Table 8 lists the D-statistics of the proposed method with varying
densification exponents and the ranking. When $d_{\alpha}$ is
greater, the sample graph created by the proposed method cannot
capture the properties of the original graph. When $d_{\alpha}$ is
smaller, the sample graph reflects the properties of the original
graph better. Thus, the closer densification exponent of the sample
graph is to that of the original graph, the more the properties of
the sample graph are similar to those of the original graph. The
results confirm that DPL-based sampling is effective, especially
when combined with community-based sampling in the proposed method.

\begin{table}[h]
\centering \caption{The performance of the proposed method with with
varying $\alpha$}
\begin{tabular}{|c|c|c|c|c|c|c|c|} \hline
$d_{\alpha}$&Degree&Sval&Svec&CC&Hop&Avg&R \\
\hline
-0.5 & 0.382 & 0.266 & 0.431 & 0.509 & 0.071 & 0.319 & 11\\
\hline

-0.4 & 0.348 & 0.191 & 0.406 & 0.466 & 0.059 & 0.294 & 10\\
\hline

-0.3 & 0.279 & 0.124 & 0.296 & 0.401 & 0.073 & 0.235 & 9\\
\hline

-0.2 & 0.231 & 0.090 & 0.296 & 0.324 & 0.068 & 0.202 & 8\\
\hline

-0.1 & 0.192 & 0.058 & 0.149 & 0.316 & 0.043 & 0.152 & 2\\
\hline

0 & 0.132 & 0.044 & 0.176 & 0.338 & 0.045 & 0.147 & 1\\
\hline

0.1 & 0.149 & 0.065 & 0.208 & 0.379 & 0.042 & 0.169 & 3\\
\hline

0.2 & 0.167 & 0.089 & 0.203 & 0.402 & 0.036 & 0.179 & 7\\
\hline

0.3 & 0.160 & 0.087 & 0.204 & 0.393 & 0.041 & 0.177 & 4\\
\hline

0.4 & 0.159 & 0.087 & 0.206 & 0.396 & 0.038 & 0.177 & 5\\
\hline

0.5 & 0.163 & 0.090 & 0.206 & 0.398 & 0.038 & 0.179 & 6\\
\hline
\end{tabular}
\end{table}
\vspace{-0.4cm}
\section{Conclusions}

In this paper, we have proposed a new sampling method that combines
two techniques: community-based sampling and DPL-based sampling.
First, it partitions the original graph into a set of sub-graphs
using hierarchical community extraction. Second, it creates sample
sub-graphs. The number of nodes and the number of edges in each
sample sub-graph are computed based on DPL. Third, it creates sample
sub-graphs by selecting nodes and edges connecting nodes in
proportion to their degree within the community. Finally, it builds
the final sample graph is created by merging the sample sub-graphs
by selecting the edges among the communities.

Through a series of experiments using a set of diverse real-world
online social networks, we have demonstrated the effectiveness of
the proposed method. The results show that the properties of the
sample graph created by our method are the most similar to those of
the original graph. We have also demonstrated the effectiveness of
the two underlying techniques, community-based sampling and
DPL-based sampling, by applying them to existing sampling methods.
The results show that community-based sampling improves the
performance of the existing sampling methods but DPL-based sampling
does not. We have shown, however, the DPL-based sampling technique
is effective when combined with community-based sampling.

\end{document}